\input{aipcheck}

\documentclass[
,final            % use final for the camera ready runs
  ]
  {aipproc}

\layoutstyle{6x9}

\begin{document}

\title{Ultra High Energy Cosmic Rays and the Auger Observatory}


\author{Antoine Letessier-Selvon for The Pierre Auger Collaboration}{
  address={Observatorio Pierre Auger, Av. San Martin Norte 304, 5613 Malarg\"ue, Argentina}
  %address={
  %Centro Brasileiro de Pesquisas Fisicas, Rua Dr. Xavier Sigaud 150,
  %22290-180  Rio de Janeiro, RJ Brasil}
}

%\author{The Pierre Auger Collaboration}{
%  address={Observatorio Pierre Auger, Av. San Martin Norte 304, 5613 Malarg\"ue, Argentina}
%}

\begin{abstract}
In this proceeding we present the construction status and the performances of the Pierre Auger Observatory together with
the first results obtained with our initial 18 month of data. In particular,
we discuss our search for anisotropy near the Galactic Center, our limit on the
photon fraction at the highest energies and our first estimate of the cosmic ray spectrum above 3 EeV.
All of the material presented in this proceeding was extracted from the numerous Auger contributions to the 29th 
ICRC proceedings.

\end{abstract}

\maketitle

%%%%%%%%%%%%%%%%%%%%%%%%%%%%%%%%%%%%%%%%%%%%
%% MAINMATTER
%%%%%%%%%%%%%%%%%%%%%%%%%%%%%%%%%%%%%%%%%%%%

\section{Introduction}
The Pierre Auger Observatory~\cite{xavier:2}  aims at unveiling the secrets of Ultra High Energy Cosmic Rays (UHECR)
through the observation of the Extensive Air Showers (EAS) they produce in the atmosphere. It combines
four fluorescence detector (FD) sites with a surface array of 1600 water Cherenkov tanks placed on a triangular 1.5 km grid. 
The combination of a large ground array and  fluorescence detectors, known as the hybrid
concept, means that a rich variety of measurements can be made on a single shower, providing much improved
information over what is possible with either detector alone. 
It is not
simply a dual experiment. Apart from important cross-checks and measurement redundancy, the two techniques
see air showers in complementary ways. The ground array measures the lateral structure of the shower
at ground level, with some ability to separate the electromagnetic and muon components. On the other hand,
the  fluorescence detector records the longitudinal profile of the shower during its development through the
atmosphere.

\section{Status and Performances of the Observatory}
\subsection{Surface Detector}%\protect\footnote{Taken from Ref.~\cite{XAVIER}}}
%(from~\cite[Xavier]}
An Auger Surface Detector (SD)  station is a 10~m$^2$ base, 1.5~m tall cylindrical plastic tank 
filled with locally produced purified water. 
Three 9" 
photo-multiplier tubes are used to collect the Cherenkov light emitted by particles crossing the tank. Signal is extracted both 
from the anode and the last dynode, the latter being amplified to achieve a larger final dynamic range extending from a few 
to about 10$^5$ photoelectrons. All channels are digitized 
at 40 MHz by 10 bit FADC, and a  digital trigger is operated by a local CPU. Timing is obtained by a GPS unit, and 
communication to the Central Data Acquisition System (CDAS) is done via a custom built wireless communication system. Two solar 
panels charging two 12 V batteries provide the 10 W used by the electronics. Each detector is therefore independent and can 
start operating upon installation, independently of other detectors in the array. 
More details about the SD can be found  in~\cite{xavier:2,XAVIER} and references therein. 
Since January 
2004, the array has been in stable operation, has grown at a steady rate of about 9 tanks per week, and reached 800
detectors in June 2005. 
Each tank is deployed and its position is verified with differential GPS technique. Even if the landscape sometimes 
forces some displacements from the perfect triangular geometry, 50~\% of the tanks are at less than 5~m from the theoretical 
position, and 90~\% at less than 20~m. The exact position is used to operate the GPS in position hold mode, achieving better 
than 20~ns time resolution.

\begin{figure}
\label{fig:xavier}
  \includegraphics[width=.90\textwidth]{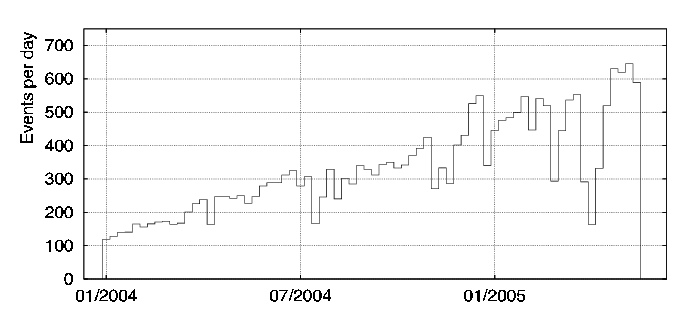}
  \caption{Evolution of event rate with time. An average event rate of about one physics event per day per station is observed.
  The consequence of a major software upgrade in April 2005 is visible as a dip in the plot. }
\end{figure}

The environment to which an Auger Surface Detector is exposed is somewhat hostile for the electronics. At 1400 m a.s.l. and 
with clear skies, day-night temperature variations are of the order of 20$^\circ$C. 
To monitor the whole array 
accurately various sensors are installed in each tank. This information is sent to the CDAS every 6 minutes. Temperature is 
measured on each PMT base, on the electronics board, and on each battery. PMT voltage and current are also monitored, as well as 
solar panel voltages, individual battery voltage, and charge current. These data are used to detect a wide range of failures, 
from broken solar panels to discharging batteries, and correlations like unstable PMT behavior related to temperature. 
Weather 
stations reporting temperature, pressure, humidity, wind speed and direction are installed at each fluorescence site and in the 
center of the array, to complete the environmental monitoring. These data allow extra checks such as the influence of the pressure 
on the calibration. The calibration is operated online every minute\cite{xavier:3}, and sent to CDAS every 6 minutes for monitoring.
Over the whole array, correlation of 
the trigger rate with temperature are -0.04$\pm$0.03\% per degree for first trigger level (T1), 0.08$\pm$0.05\% per degree for the 
second level trigger (T2), and 0.2$\pm$0.5\% per degree for the Time over Threshold trigger (ToT). The SD array therefore operates 
with stable trigger threshold even with 20 degrees daily temperature variations. 

The last step monitored to ensure the quality of the Auger SD data set is done at the system acquisition level. The second 
level trigger rate for each station are registered every second allowing a precise knowledge of the dead time of the detectors. 
The acquisition is fully automated and no operator is needed for data taking.  Information from the CDAS processes are kept to 
diagnose possible crashes. 
Simple quantities such as the number of stations in operation and the event rate, and more complex ones such as the rate of physics
events are checked daily to validate the data acquisition period. Over 2004, the total on-time of the system has been about 94~\%, 
including all kinds of dead time (individual detectors down, general power cuts, software upgrades, etc.).  It should be noted that
this on-time was obtained while priority was being given to the building of the Observatory (deploying new detectors) over its 
operation (repairing failing ones), and with evolving software for the detectors, the communication system, and the CDAS.

Up to June 2005, more than 180000 events were recorded with an average rate of about 0.9 per station per day 
(see Fig.~\ref{fig:xavier}). Once the array is completed, a rate of about 1500 physics events per day is expected.

\subsection{Fluorescence Detector}%\protect\footnote{Taken from Ref.~\cite{JOSE}}}
The fluorescence detectors are distributed in 4 stations around the perimeter of the surface detector array, and 
view the atmosphere above the array on moon-less or partially moon-lit nights. At the present time three of the four 
fluorescence sites have been completed and are in operation. 
Two of them, Los Leones and Coihueco, have been collecting data since January 2004, with Los Morados beginning data taking in 
March 2005. The fourth site at Loma Amarilla will be in operation in the second half of 2006. A fluorescence site contains 
six identical fluorescence telescopes. Fluorescence light enters the telescope
through a 1.10~m radius diaphragm, and light is collected by a 3.5x3.5~m$^2$ spherical mirror and focused onto a photo-multiplier
(PMT) camera. The camera contains 440 hexagonal (45~mm diameter) PMTs, each PMT covering a 1.5$^\circ$ diameter portion of the 
sky. The optical spot size on the focal surface has a diameter of approximately 15~mm (equivalent to 0.5$^\circ$) for all
directions of incoming light. To reduce signal losses when the light spot crosses PMT boundaries, small light reflectors 
("Mercedes stars") are placed between PMTs. The field of view of a single telescope covers 30$^\circ$ in azimuth and 28.6$^\circ$
in elevation. 
The fluorescence telescopes have been installed with an uncertainty of 0.1$^\circ$ in their nominal pointing directions. 
However, observations of stars crossing the field of view of the telescopes can improve this precision, to 0.01$^\circ$. 
An optical filter matched to the fluorescence spectrum (approximately 300~nm to 400~nm) is placed over the telescope 
diaphragm to reduce night-sky noise. In addition, the diaphragm contains an annular corrector lens as part of the Schmidt 
telescope design, with an inner radius of 0.85~m and outer radius of 1.10~m. The effect of the lens is to allow an increase 
in the radius of the telescope diaphragm from 0.85~m to 1.1~m (increasing the effective light collecting area by a factor of 
two) while maintaining an optical spot size of 0.5$^\circ$~\cite{jose:4}. 

One of the goals of the FD is to measure air shower energies with an uncertainty smaller than 15\%. In order to achieve this 
goal the fluorescence detectors have to be calibrated with a precision of about 8\% and the calibration stability needs to 
be monitored on a regular basis. An absolute calibration of each telescope is performed three or four times a year, and 
relative calibrations are performed every night during detector operation. To perform an absolute end-to-end calibration of 
a telescope, a large homogeneous diffuse light source was constructed for use at the front of the telescope diaphragm. 
The ratio of the light source intensity to the observed signal for each PMT 
gives the required calibration. At present, the precision in the PMT calibration using the source is about 12\%~\cite{jose:5}. 
For relative calibration, 
optical fibers bring light signals to three different diffuser groups for each telescope
The total charge per pixel is measured with respect to reference
measurements made at the time of absolute calibrations. This allows the monitoring of short and long term stability, 
the relative timing between pixels and the relative gain of each pixel~\cite{jose:6}. The relative calibration information is not 
yet incorporated in the reconstruction system. However, the average detector stability has been measured and a corresponding
systematic uncertainty of 3\% has been introduced to account for this. This contributes to the overall 12\% systematic 
uncertainty in the FD calibration. Cross-checks of the FD calibration can be made by reconstructing the energy of laser beams 
that are fired into the atmosphere from various positions in the SD array. The Central Laset Facility (CLF see next section) 
located at the center of the array allows to fire laser beam into  to the sky with known geometry 
and energy. 
The observed difference between the reconstructed energy 
and the real laser energy is of the order of 10\% to 15\%~\cite{jose:8}, consistent with the current level of uncertainty 
in calibrations and knowledge of the atmosphere. 

As part of the reconstruction process, the detected light at the telescope must be transformed into the amount of 
fluorescence light emitted at the shower axis as a function of atmospheric depth. For this it is necessary to have a good 
knowledge of local atmospheric conditions. We need to account for both Rayleigh and aerosol scattering of light between the 
shower and the detector, so we must understand the distribution of aerosols and the density of the atmosphere at different 
heights. In addition, the temperature distribution with height is needed since the fluorescence light yield is a (slow) 
function of temperature. Finally, the detector volume must be monitored for the presence of clouds. Aerosols in the atmosphere
consist of clouds, smoke, dust and other pollutants. The aerosol conditions can change rapidly and are known to have a strong
effect on the propagation of fluorescence light. The Observatory has an extensive network of atmospheric monitoring devices. 
These include LIDAR systems, cloud cameras and star monitors. We have also deployed systems to monitor the wavelength dependence
and differential scattering properties of the aerosols. More details of these systems can be found in~\cite{jose:7}. 
Presently, only the aerosol information obtained from observing the laser tracks is incorporated in the shower energy
reconstruction algorithm. 
The uncertainty in the 
currently applied monthly atmospheres in the Auger reconstruction introduce an uncertainty in the atmospheric depth at ground 
of about 5 g/cm$^2$\cite{jose:10}. 

The resulting fluorescence light at the shower track is converted to the energy 
deposited by the shower by applying the expected fluorescence efficiency at each depth. 
More details about the FD calibration and performances  can be found  in~\cite{xavier:2,JOSE} and references therein. 
The estimated systematic uncertainty 
in the reconstructed shower energy is currently 25\%, with activity underway to reduce this significantly.

\begin{figure}
\label{fig:miguel}
  \includegraphics[width=.90\textwidth]{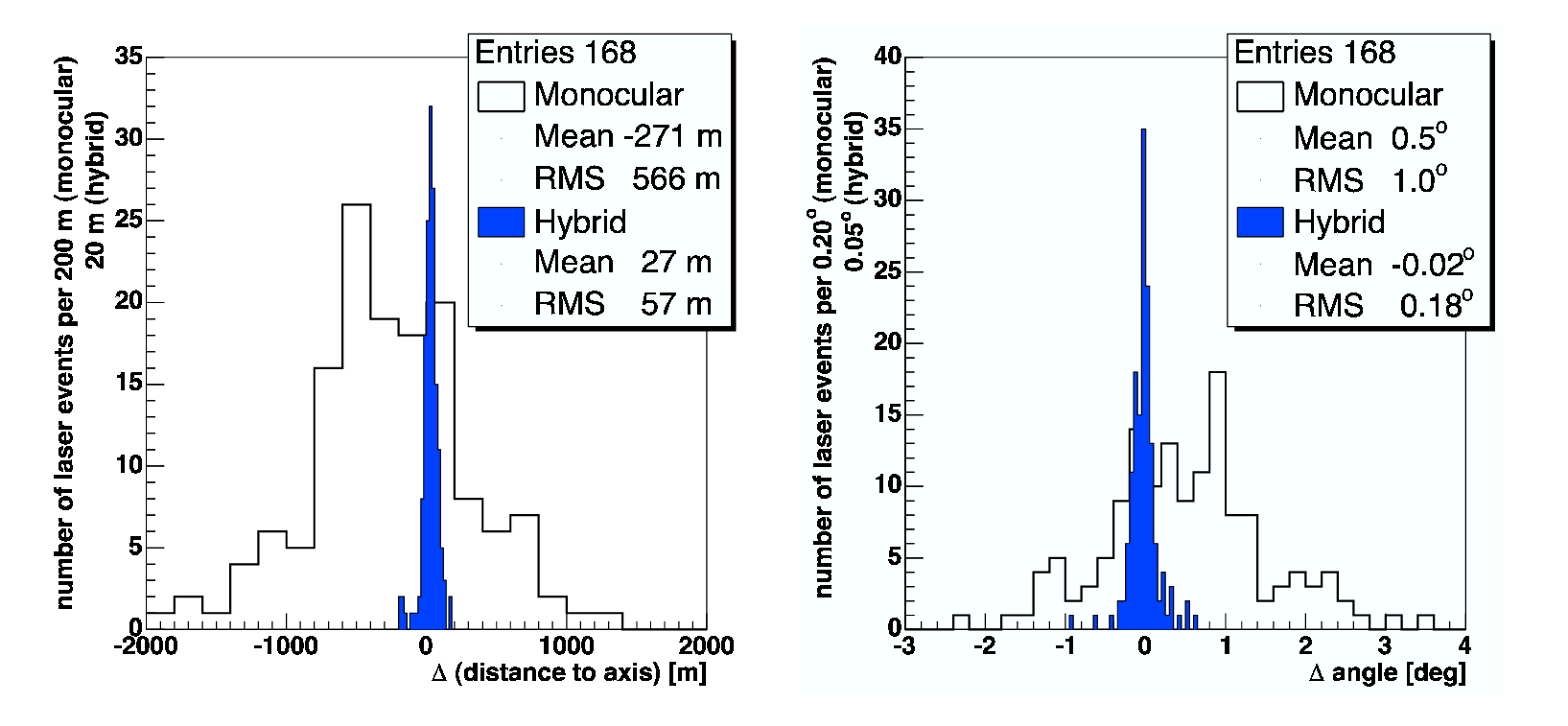}
  \caption{Left : Difference between the reconstructed and true distance from the eye to the vertical laser beam using the
  monocular and hybrid techniques. The location of the laser is known to 5 m. Right :  Angular difference between reconstructed 
  and true direction of the laser beam using the monocular and hybrid techniques. The laser beam is vertical within 0.01$^\circ$. }
\end{figure}

\subsection{Hybrid Performances}%\protect\footnote{Taken from Ref.~\cite{MIGUEL}}}
A hybrid event is an air shower that is simultaneously detected by the fluorescence detector and the ground array. 
The Observatory was originally designed and is currently being built with a cross-triggering capability. Data are recovered 
from both detectors whenever either system is triggered. If an air shower independently triggers both detectors the event is 
tagged accordingly. There are cases where the fluorescence detector, having a lower energy threshold, promotes a sub-threshold 
array trigger. Surface stations are matched by timing and location. This is an important capability because these sub-threshold 
hybrid events would not have triggered the array otherwise. The Observatory started operation in hybrid production mode in January,
2004. Surface stations have a 100\% duty cycle, while fluorescence eyes can only operate on clear moon-less nights. 
Both surface and fluorescence detectors have been running simultaneously 14\% of the time. The number of hybrid events represents 
10\% the statistics of the surface array data. 
 
A hybrid detector has excellent capability for studying the highest energy cosmic ray air showers. Much of its capability stems 
from the accurate geometrical reconstructions it achieves. Timing information from even one surface station can much improve 
the geometrical reconstruction of a shower over that achieved using only eye pixel information. The axis of the air shower is 
determined by minimizing a $\chi^2$  function involving data from all triggered elements in the eye and at ground. The 
reconstruction accuracy is better than the ground array counters or the single eye could achieve 
independently~\cite{miguel:1,miguel:2}. 
Using the timing information from the eye pixels together with the surface stations, a core location resolution of 50~m is 
achieved. The resolution for the arrival direction of cosmic rays is 0.6$^\circ$~\cite{miguel:2}. 
These results for the hybrid accuracy are in 
good agreement with estimations using analytic arguments~\cite{miguel:3}, measurements on real data using a bootstrap method\cite{miguel:4},
and previous 
simulation studies~\cite{miguel:5}. The reconstruction uncertainties are evaluated using events with known geometries, i.e. laser beams. The 
CLF, described in Ref.~\cite{jose:8}, is located approximately equidistant from the first three 
fluorescence sites. Since the location of the CLF and the direction of the laser beam are known to an accuracy better than the 
expected angular resolution of the fluorescence detector, laser shots from the CLF can be used to measure the accuracy of the 
geometrical reconstruction. Furthermore, the laser beam is split and part of the laser light is sent through an optical fiber 
to a nearby ground array station. The resolution of the monocular and hybrid reconstructions are compared in 
figure~\ref{fig:miguel} for the  distance between the eye and the CLF, and for the angle of the axis. 

The laser light from the CLF produces simultaneous triggers in both the surface and (three) fluorescence detectors. The 
recorded event times can be used to measure and monitor the relative timing between the two detectors. The time offset between 
the first fluorescence eye and the surface detector is shown in figure 3. This time offset has been measured to better than 
50~ns~\cite{miguel:7}. The contribution to the systematic uncertainty in the core 
location due to the uncertainty in the time synchronization
is 20~m. 
More details about the Hybrid performances  of the Auger Observatory can be found  in~\cite{xavier:2,MIGUEL} and references therein. 

Due to the much improved angular accuracy, the hybrid data sample is ideal for anisotropy studies.
Many ground parameters,like the shower front curvature and thickness, have always been difficult to measure experimentally,
and were usually determined from Monte Carlos simulation. The hybrid sample provides a unique opportunity in this respect.
As mentioned, the geometrical reconstruction can be done using only one ground station,thus all the remaining detectors can be 
used to measure the shower characteristics.  The possibility of studying the same set of air showers with two independent methods
is valuable in understanding the strengths and limitations of each technique. The hybrid analysis benefits from the calorimetry 
of the fluorescence technique and the uniformity of the surface detector aperture. 

\section{Results Highlights}
\subsection{Anisotropy Studies Around the Galactic Center}%\protect\footnote{Taken from Ref.~\cite{ANTOINE}}}

The galactic centre (GC) region provides an attractive target for
anisotropy studies with the Pierre Auger Observatory. On the one hand,
there have been in the past observations by the
AGASA~\cite{antoine:1} and SUGAR~\cite{antoine:2} experiments indicating an excess of cosmic rays from this region in the EeV energy range . On the other hand, since the GC harbors a very massive black hole,
it provides a natural candidate for CR accelerator to very high energies.

In this study Auger data from 1$^{st}$ January 2004 until 6$^{th}$ June 2005 was used.
Events from the surface detector that passed the 3-fold or
the 4-fold data acquisition triggers and satisfying our high level physics trigger (T4)
and our quality trigger (T5)~\cite{antoine:14} were selected.
The T5 selection is
independent of energy and ensures a better quality for the event reconstruction.
This data set has an angular resolution better than 2.2$^\circ$ for all of the 3-fold
events (regardless of the zenith angle considered) and better than 1.7$^\circ$ for all events with
multiplicities > 3 SD stations~\cite{miguel:2}. In all our analyses the zenith angle was cut at
60$^\circ$ like AGASA while SUGAR used all zenith angles.
\begin{figure}[t]
\includegraphics[width=.90\textwidth]{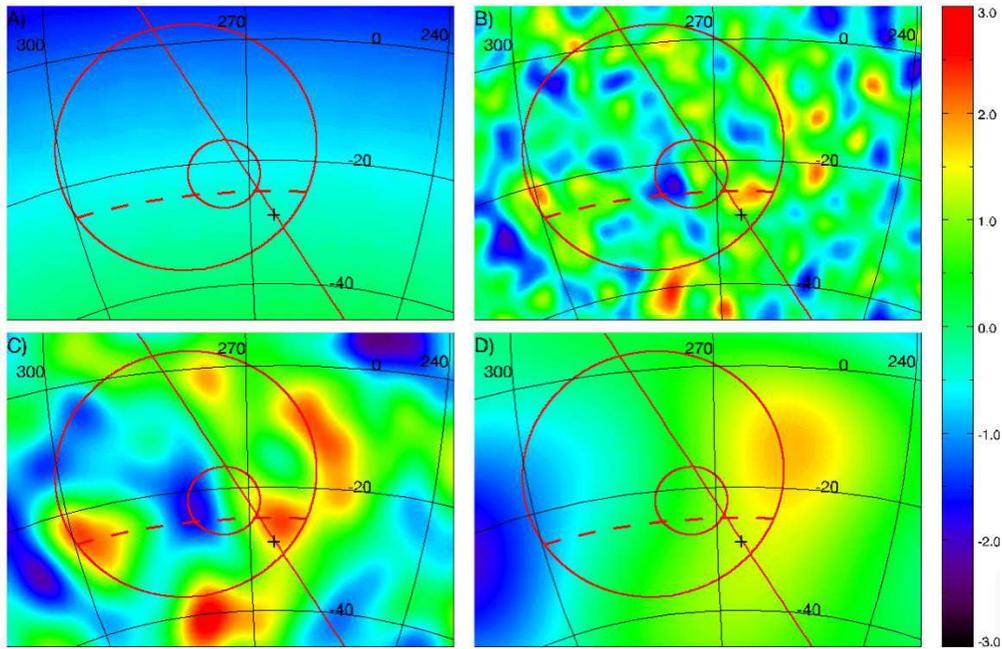}
%\end{center}
\vspace*{-0.5cm}
\caption{Lambert projections of the galactic centre region, GC (cross), galactic plane (solid line),  regions of excess of AGASA and SUGAR (circles), AGASA f.o.v. limit (dashed line).
A) coverage map (same color scale as the significance maps, but in a range [0-1.0]).
B) significance map in the range [0.8-3.2]~EeV smoothed using the individual pointing resolution of the events and
a 1.5$^\circ$ filter (Auger like excess),
C) same  smoothed at 3.7$^\circ$ (SUGAR like excess), D) in the range [1.0-2.5]~EeV smoothed at 13.3$^\circ$ (AGASA like excess).}
\label{fig:1}
\end{figure}

To estimate the coverage map, needed to construct excess and excess probability maps, a shuffling technique was used.
In Fig.~3A the coverage map obtained from our SD sample in a region around the GC is presented. 
In Fig.~3 B,C and D we present the chance probability distributions (mapped to positive Gaussian significance for excesses and
negative for deficits) in the same region for various filtering and energy cuts corresponding to our various searches.
In these maps the chance probability distributions are consistent with those expected as a result of statistical
fluctuations from an isotropic sky.

Regarding the region where the AGASA excess was reported, the results from the Auger Observatory are $1155$ events observed,
and $1160.7$ expected (ratio 1.00$\pm$0.03) for the energy range [1.0-2.5]~EeV.
These results do not support the excess observed by AGASA, and
in particular not at a level of 22\% like the one they reported which would translate into
a 7.5$\sigma$ excess. In a worst case scenario where the source would be protons and the background much heavier (e.g. Iron), the
difference in detection efficiency of the Auger trigger at 1~EeV would reduce the sensitivity to a source excess. However,
using the Fe/proton efficiency ratio at 1~EeV ($70\%/50\%=1.44$, an upper bound in the range [1-2.5]~EeV)
a 5.2$\sigma$ event excess would still be expected in our data set.

Regarding the excess claimed  by SUGAR, we find in their angular/energy window
$144$ events observed, and $150.9$  expected (ratio 0.95$\pm$0.08)
, and hence with over an order of magnitude more statistics we are not able to confirm this claim.

A search was performed for signals of a point-like source in the direction
of the GC. Using a 1.5$^\circ$ Gaussian filter corresponding to the angular resolution of the SD~\cite{miguel:2}.
In the energy range
[0.8--3.2]~EeV, we obtain $24.3$ events observed and, $23.9$ expected (ratio 1.0$\pm$0.1).
A  95\% CL upper bound on  the number of events coming from a point source in that window is
$n_s(95\%)= 6.7$. This bound can be translated into a
flux upper limit ($\Phi_s$) integrated in this energy range.
In the simplest case in which the source has a spectrum
similar to the one of the overall CR spectrum (d$N/{\rm d}E\propto E^{-3}$),
$\Phi_s = n_s \Phi_{CR} 4\pi\sigma^2/n_{exp}$
where $\sigma$ is the size of the Gaussian filter used.
Using $\Phi_{CR}(E)= 1.5\ \xi (E/EeV)^{-3} \times 10^{-12}\, (\mbox{\rm EeV$^{-1}$\,m$^{-2}$\,s$^{-1}$\,sr$^{-1}$})$
where $\xi \in [1,2.5]$ denotes our uncertainty on the CR flux ($\xi$ is around unity for Auger and 2.5 for AGASA),
introducing $\varepsilon$ the Iron/proton detection  efficiency ratio ($1< \varepsilon < 1.6$ for $ E \in [0.8,3.2]$~EeV) and,
integrating in that energy range we obtain :
%\[
$$
\Phi_s  < 2.6\,\, \xi \,\, \varepsilon \times 10^{-15}\, \mbox{\rm m$^{-2}$s$^{-1}$} \mbox{~~~~@ 95\% CL.}
$$
%\]

In a worst case scenario, where both $\xi$ and $\varepsilon$
take their maximum value, the bound is $\Phi_s = 10.6 \times 10^{-15}\, \mbox{\rm m$^{-2}$s$^{-1}$}$, and
still excludes the neutron source scenario suggested in~\cite{antoine:1,antoine:9} to
account for the AGASA excess,  or in~\cite{antoine:3,antoine:4} in connection with the HESS measurements.
More details about the GC anisotropy studies with the Auger Observatory data can be found  in~\cite{ANTOINE}.

\subsection{A First Estimate of the Cosmic Ray Spectrum Above 3~EeV}%\protect\footnote{Taken from Ref.~\cite{PAUL}}}
The data for this analysis are from 1 Jan 2004 through 5 Jun 2005.
The event acceptance criteria and exposure calculation are described
in separate papers \cite{antoine:14,paul:6}.  Events are included
for zenith angles 0-60$^{\circ}$, and results are reported for
energies above 3 EeV (3525 events). The array is fully efficient for
detecting such showers, so the acceptance at any time is the simple
geometric aperture.  The cumulative exposure adds up to 1750 km$^2$ sr
yr, which is 7\% greater than the total exposure obtained by AGASA
\cite{AGASA}.  The average array size during the time of this exposure
was 22\% of what will be available when the southern site of the
Observatory has been completed.

Assigning energies to the SD event set is a two-step process.
The first step is to assign an energy parameter $S_{38}$ to each
event.  Then the hybrid events are used to establish
the rule for converting $S_{38}$ to energy.
The energy parameter $S_{38}$ for each shower comes from its
experimentally measured S(1000), which is the time-integrated
water Cherenkov signal S(1000) that would be measured by a tank
1000 meters from the core.  

The signal S(1000) is
attenuated at large slant depths.  Its dependence on zenith angle
is derived empirically by exploiting the nearly isotropic
intensity of cosmic rays.  By fixing a specific intensity $I_0$
(counts per unit of $sin^2\theta$), one finds for each zenith
angle the value of S(1000) such that $I(>S(1000))=I_0$.  We
calculated a particular constant intensity cut curve $CIC(\theta)$
relative to the value
at the median zenith angle ($\theta\approx 38^{\circ}$).  Given
S(1000) and $\theta$ for any measured shower, the energy
parameter $S_{38}$ is defined by {\bf $S_{38}\equiv
S(1000)/CIC(\theta)$}.  It may be regarded as the S(1000)
measurement the shower would have produced if it had arrived
$38^{\circ}$ from the zenith.

$S_{38}$ is well correlated with the FD energy measurements in
hybrid events that are reconstructed independently by the FD and
SD.  A linear relation was fitted and gives an empirical rule for
assigning energies (in EeV) based on $S_{38}$ (in VEM):
\begin{equation}E = 0.16 \times S_{38}^{1.06} = 
  0.16\times [S(1000)/CIC(\theta)]^{1.06}.\end{equation}
The uncertainty in this rule is discussed below.  

\begin{figure}[t]
\begin{minipage}[t]{0.5\linewidth}
\includegraphics*[width=1.0\textwidth]{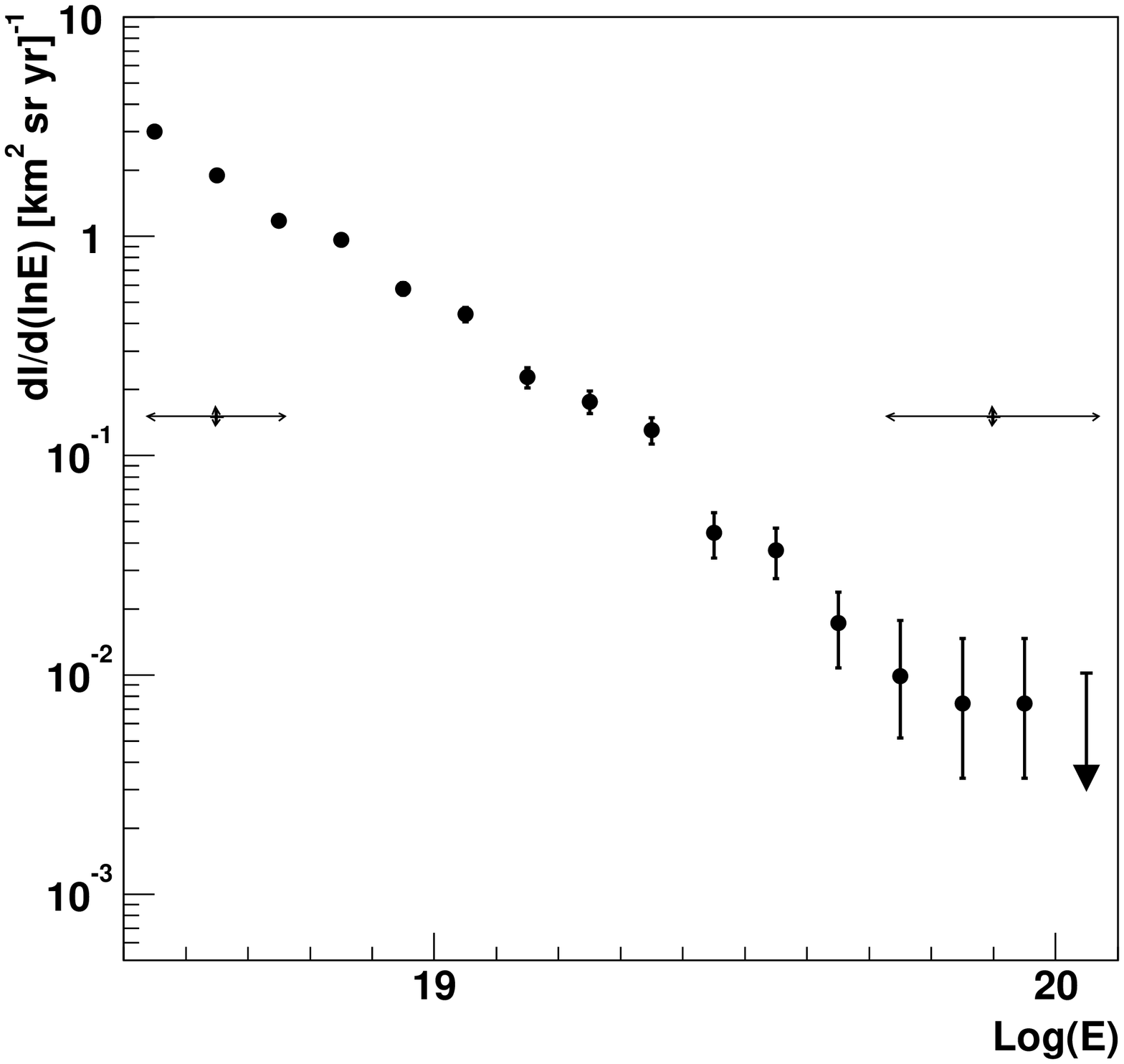}
\end{minipage}
\hfill
\begin{minipage}[t]{0.5\linewidth}
\includegraphics*[width=1.0\textwidth]{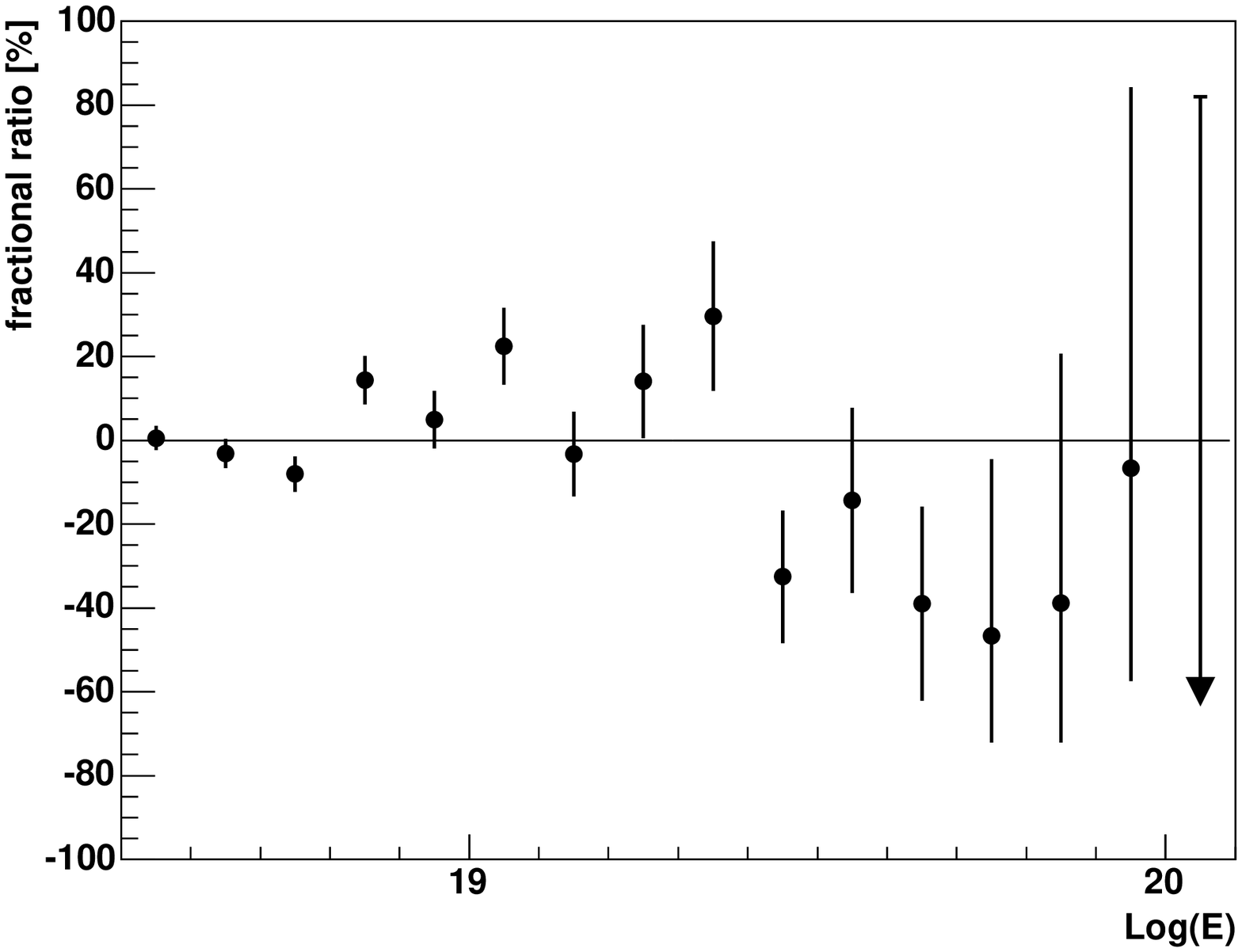}
\caption{\label{fig:paul} Left : Estimated spectrum.  Plotted on the vertical axis is
the differential intensity $\frac{dI}{dlnE}\equiv
E\frac{dI}{dE}.$  Error bars
on points indicate statistical
uncertainty (or 95\% CL upper limit).  Systematic uncertainty is indicated by double
arrows at two different energies. Right: Percentage deviation from the best-fit
power law: $100\times((dI/d(lnE)-F)/F$.  The fitted function is $F =
30.9\pm 1.7\times(E/EeV)^{-1.84\pm 0.03}$.  The chisquare per degree of freedom
in the fit is 2.4}
\end{minipage}
\hfill
\end{figure}

The distribution over $ln(E)$ produced by this two-step procedure
becomes the energy spectrum of figures~\ref{fig:paul} after dividing by the
exposure: 1750 km$^2$ sr yr.  (See also
http://www.auger.org/icrc2005/spectrum.html.)

The Auger Observatory will measure the spectrum over the southern
sky accurately in coming years.  The spectrum in figure~\ref{fig:paul} is only
a first estimate.  It has significant systematic and statistical
uncertainties.  The indicated statistical error for each point
comes directly from the Poisson uncertainty in the number of
measured showers in that logarithmic energy bin.  Systematic and
statistical uncertainties in S(1000) are discussed elsewhere~\cite{paul:7}.  
There is larger systematic uncertainty in the
conversion of $S_{38}$ to energy.  Part of that comes from the FD
energies themselves.  Laboratory measurements of the fluorescence
yield are uncertain by 15\%, and the absolute calibration of the
FD telescopes is presently uncertain by 12\%.  Together with
other smaller FD uncertainties, the total systematic uncertainty
in the FD energy measurements is estimated to be 25\%.  
Combining in quadrature the FD systematic
uncertainty and this correlation uncertainty, the total
systematic energy uncertainty grows from 30\% at 3 EeV to 50\% at
100 EeV.  This uncertainty is indicated by horizontal double
arrows in figure~\ref{fig:paul}, and a 10\% systematic uncertainty in the
exposure is indicated by vertical arrows.
More details about this analysis can be found  in~\cite{PAUL} and references therein. 

The Pierre Auger Observatory is still under construction and
growing rapidly.  By the next ICRC meeting, its cumulative
exposure will be approximately 7 times greater.  The statistical
errors will shrink accordingly, permitting a search in the
southern skies for spectral features, including the predicted GZK
suppression.  The enlarged hybrid data set will reduce systematic
uncertainty in the FD normalization of the SD energies.
Numerous laboratory experiments are attempting to reduce the
systematic uncertainty in the fluorescence yield, which will
be the dominant uncertainty in the FD normalization of the
Auger energy spectrum.  The FD detector calibration uncertainty
will also be reduced.

\subsection{An Upper Limit on the Primary Photon Fraction}%\protect\footnote{Taken from Ref.~\cite{MARKUS}}}
The photon upper limit derived
here is based on the direct observation of the longitudinal air
shower profile and makes use of the hybrid detection technique:
$X_{\rm max}$ is used as discriminant observable.
The information from triggered surface detectors in hybrid
events considerably reduces the uncertainty in shower track geometry.

\begin{figure}[t]
\noindent
\begin{minipage}[l]{.5\linewidth}
\includegraphics[width=.99\textwidth]{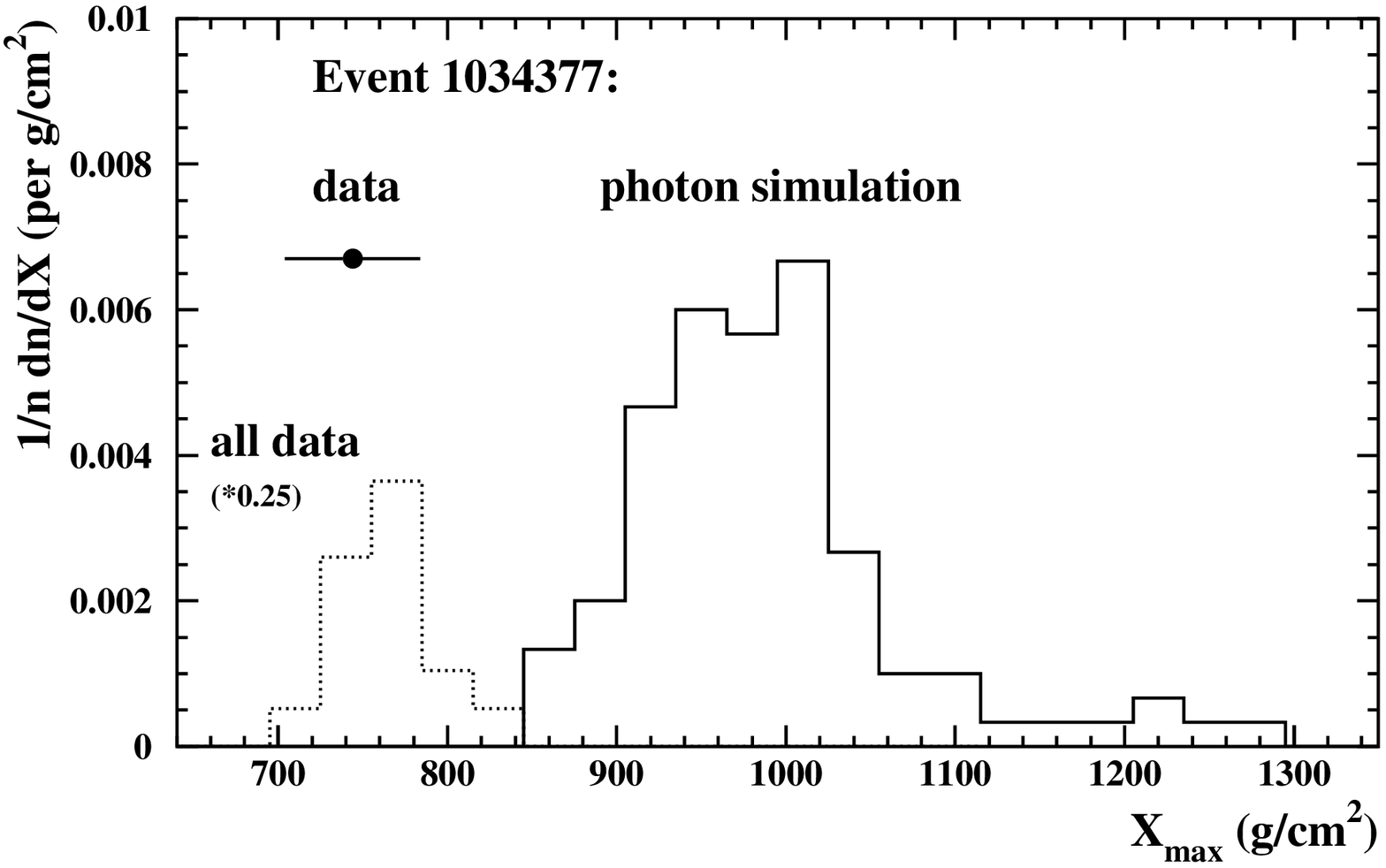}
\end{minipage}\hfill
\begin{minipage}[c]{.5\linewidth}
\includegraphics[width=.99\textwidth]{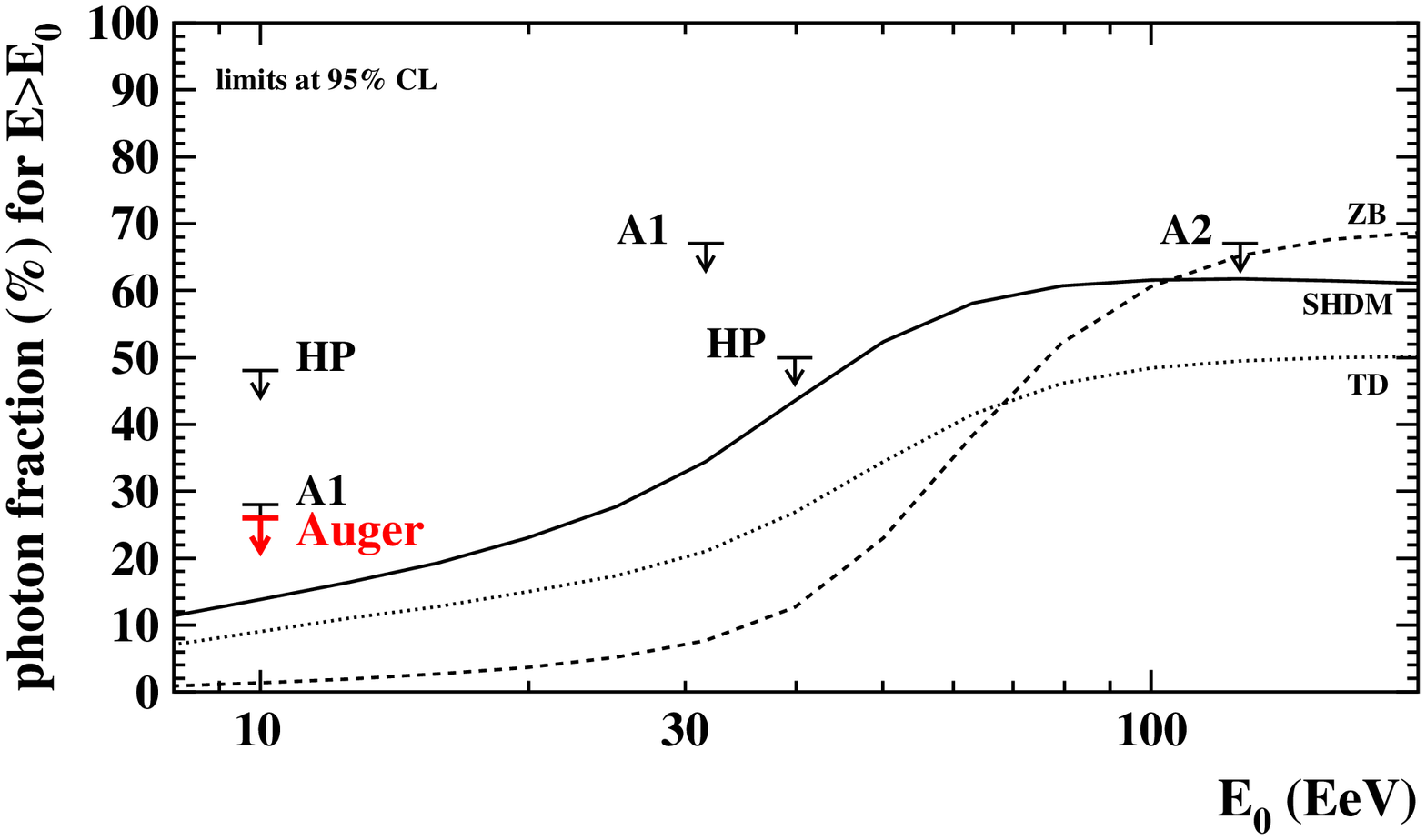}
\caption{Left :Example of $X_{\rm max}$ measured in an individual shower
of 11~EeV (point
with error bar) compared to the $X_{\rm max}$ distribution
expected for photon showers (solid line).
Also shown the $X_{\rm max}$ distribution of the data sample (dashed line;
normalization changed as indicated).
Right: Upper limits (95\% CL) on cosmic-ray photon fraction
derived in the present analysis (Auger) and
previously from AGASA (A1)~\cite{shinozaki}, (A2)~\cite{risse05}
and Haverah Park (HP)~\cite{ave} data compared to some estimates
based on non-acceleration models~\cite{models}.}
\label{fig:markus}
\end{minipage}
\end{figure}

The data are taken with a total of 12 fluorescence
telescopes~\cite{JOSE}, situated at
two different telescope sites, during the period January 2004
to April 2005. The number of deployed surface detector
stations ~\cite{XAVIER} grew
from $\sim$200 to $\sim$800 during this time.
For the analysis, hybrid events were selected, i.e.~showers observed
both by (at least one) surface tank and telescope~\cite{MIGUEL}.
Even for one triggered tank only, the additional timing constraint allows
a significantly improved geometry fit to the observed profile
which leads to a reduced uncertainty in the reconstructed $X_{\rm max}$.

The reconstruction is based on an end-to-end calibration of the
fluorescence telescopes~\cite{brack}, on monitoring data of
local atmospheric conditions~\cite{keilhauer,jose:7},
and includes an improved subtraction of Cherenkov
light~\cite{nerling}
and reconstruction of energy deposit profiles for deriving the
primary energy.
In total, 16 events with energies above $10^{19}$~eV  are selected.

The total uncertainty $\Delta X_{\rm max}^{\rm tot}$ of the
reconstructed depth of shower maximum is composed of
several contributions which, in general, vary from event to event.
A conservative estimate of the current $X_{\rm max}$ uncertainties
gives $\Delta X_{\rm max}^{\rm tot}\simeq$ 40~g~cm$^{-2}$.
Among the main contributions, each one in general well below
$\Delta X_{\rm max}=$15~g~cm$^{-2}$, are
the statistical uncertainty from the profile fit,
the uncertainty in shower geometry,
the uncertainty in atmospheric conditions such as the air
density profile, and
the uncertainty in the reconstructed primary energy, which is taken
as input for the primary photon simulation.

For each event, high-statistics shower simulations are performed for
photons for the specific event conditions.
A simulation study of the detector acceptance to photons and nuclear
primaries has been conducted.
For the chosen cuts, the ratio of the acceptance to photon-induced showers
to that of nuclear primaries (proton or iron nuclei) is $\epsilon = 0.88$.
A corresponding correction is applied to the derived photon limit.

Fig.~\ref{fig:markus} shows as an example an event of 11~EeV primary energy
observed with
$X_{\rm max} = 744$~g~cm$^{-2}$, compared to the corresponding
$X_{\rm max}$ distribution expected for primary photons.
With $<$$X_{\rm max}^\gamma$$> = 1020$~g~cm$^{-2}$, photon showers are on
average expected to reach maximum at depths considerably greater than
observed.
Shower-to-shower fluctuations are large due to the LPM effect
(rms of 80~g~cm$^{-2}$) and well in excess of the measurement
uncertainty.
For all 16 events, the observed $X_{\rm max}$ is well below the average value
expected for photons.
The $X_{\rm max}$ distribution of the data is also displayed in Fig.~\ref{fig:markus}.
More details about this analysis can be found  in~\cite{MARKUS}. 

The statistical method for deriving an upper limit follows that
introduced in~\cite{risse05}.
For the Auger data sample, an upper limit on the
photon fraction of 26\% at a confidence level of 95\% is derived.
In Fig.~\ref{fig:markus}, this upper limit is plotted together with previous experimental
limits and some estimates based on non-acceleration models.
The presented 26\% limit confirms and improves the existing limits
above $10^{19}$~eV.

\section{Prospects}
It is important to note that the Pierre Auger Observatory is under construction and that results are preliminary. 
Growing rapidly, its cumulative exposure will be approximately 7 times greater than toady within two years (mid 2007) from now.
The statistical errors on our results 
will shrink accordingly, permitting a search in the southern skies for spectral features, 
including the predicted GZK suppression, cosmic rays sources as well as primary identification. 

It is already clear that the
combination of fluorescence and ground array measurements provides reconstruction of the geometry of the
shower with much greater accuracy than is achieved with either detector system on its own. Unprecedented
core location and direction precision leads to excellent energy and shower development measurements. 
The enlarged hybrid data set will also reduce systematic uncertainty in the FD normalization of the SD energies.


\bibliographystyle{aipproc}   % if natbib is available

%% The following lines show an example how to produce a bibliography
%% without the help of the BibTeX program. This could be used instead
%% of the above.
%%%%%%%%%%%%%%%%%%%%%%%%%%%%%%%%%%%%%%%%%%%

\end{document}